\documentclass[conference]{IEEEtran}
\IEEEoverridecommandlockouts
\usepackage{cite}
\usepackage{amsmath,amssymb,amsfonts}
\usepackage{algorithmic}
\usepackage{graphicx}
\usepackage{textcomp}
\usepackage{xcolor}
\def\BibTeX{{\rm B\kern-.05em{\sc i\kern-.025em b}\kern-.08em
    T\kern-.1667em\lower.7ex\hbox{E}\kern-.125emX}}
    
\usepackage{hyperref}
\usepackage{multirow}
\usepackage{hyperref}
\newcommand\Tstrut{\rule{0pt}{2.0ex}}         
\newcommand\Bstrut{\rule[-0.9ex]{0pt}{0pt}} 
\usepackage{amssymb}

\begin{document}
\title{Dual-Domain Coarse-to-Fine Progressive Estimation Network for Simultaneous Denoising, Limited-View Reconstruction, and Attenuation Correction of Cardiac SPECT}

\author{%
  \IEEEauthorblockN{%
    \parbox{\linewidth}{\centering
        Xiongchao Chen$^{1}$\IEEEauthorrefmark{2},
        Bo Zhou$^{1}$,
        Xueqi Guo$^{1}$,
        Huidong Xie$^{1}$,
        Qiong Liu$^{1}$,
        James S. Duncan$^{1,2}$,
        Albert J. Sinusas$^{1,2,3}$,
        Chi Liu$^{1,2}$
    }%
  }%
\IEEEauthorblockA{%
    $^{1}$Department of Biomedical Engineering, Yale University, New Haven, CT 06511 US \\ 
    $^{2}$Department of Radiology and Biomedical Imaging, Yale University, New Haven, CT 06510 US \\
    $^{3}$Department of Internal Medicine, Yale University, New Haven, CT 06510 US \\
    Correspondence: \IEEEauthorrefmark{2}xiongchao.chen@yale.edu
    }%
}
\maketitle
\thispagestyle{plain}
\pagestyle{plain}

\begin{abstract}
Single-Photon Emission Computed Tomography (SPECT) is widely applied for the diagnosis of coronary artery diseases. Low-dose (LD) SPECT aims to minimize radiation exposure but leads to increased image noise. Limited-view (LV) SPECT, such as the latest GE MyoSPECT ES system, enables accelerated scanning and reduces hardware expenses but degrades reconstruction accuracy. Additionally, Computed Tomography (CT) is commonly used to derive attenuation maps ($\mu$-maps) for attenuation correction (AC) of cardiac SPECT, but it will introduce additional radiation exposure and SPECT-CT misalignments. Although various methods have been developed to solely focus on LD denoising, LV reconstruction, or CT-free AC in SPECT, the solution for simultaneously addressing these tasks remains challenging and under-explored. Furthermore, it is essential to explore the potential of fusing cross-domain and cross-modality information across these interrelated tasks to further enhance the accuracy of each task. Thus, we propose a Dual-Domain Coarse-to-Fine Progressive Network (DuDoCFNet), a multi-task learning method for simultaneous LD denoising, LV reconstruction, and CT-free $\mu$-map generation of cardiac SPECT. Paired dual-domain networks in DuDoCFNet are cascaded using a multi-layer fusion mechanism for cross-domain and cross-modality feature fusion. Two-stage progressive learning strategies are applied in both projection and image domains to achieve coarse-to-fine estimations of SPECT projections and CT-derived $\mu$-maps. Our experiments demonstrate DuDoCFNet's superior accuracy in estimating projections, generating $\mu$-maps, and AC reconstructions compared to existing single- or multi-task learning methods, under various iterations and LD levels. The source code of this work is available at \href{https://github.com/XiongchaoChen/DuDoCFNet-MultiTask}{\color{blue}{https://github.com/XiongchaoChen/DuDoCFNet-MultiTask}}.
\end{abstract}
\begin{IEEEkeywords}
Cardiac SPECT, denoising, limited-view reconstruction, attenuation correction, dual-domain learning, multi-task learning
\end{IEEEkeywords}
%
\section{Introduction}
\label{sec:introduction}
Myocardial perfusion imaging (MPI) using Single-Photon Emission Computed Tomography (SPECT) is the most widely performed nuclear medicine exam for the diagnosis of coronary artery diseases. Reducing the injected dose can lower the potential risk of radiation to patients, but it will increase the image noise \cite{einstein2012effects}. Acquiring limited-view (LV) projections using fewer solid-state detectors enables accelerated scanning and reduces hardware costs. However, it can lead to lower reconstruction accuracy due to reduced angular sampling \cite{zhu2013improved}. Additionally, Computed Tomography (CT)-derived attenuation maps ($\mu$-maps) are commonly employed for attenuation correction (AC) to improve the diagnostic accuracy of cardiac SPECT \cite{goetze2007attenuation}. However, the extra CT scans will cause additional radiation exposure and introduce SPECT-CT misalignments \cite{saleki2019influence}. Additionally, stand-alone SPECT scanners without CT assistance dominate the SPECT market share, where the conventional CT-based AC approach is not available.

\begin{figure*}[htb!]
\centering
\includegraphics[width=0.92\textwidth]{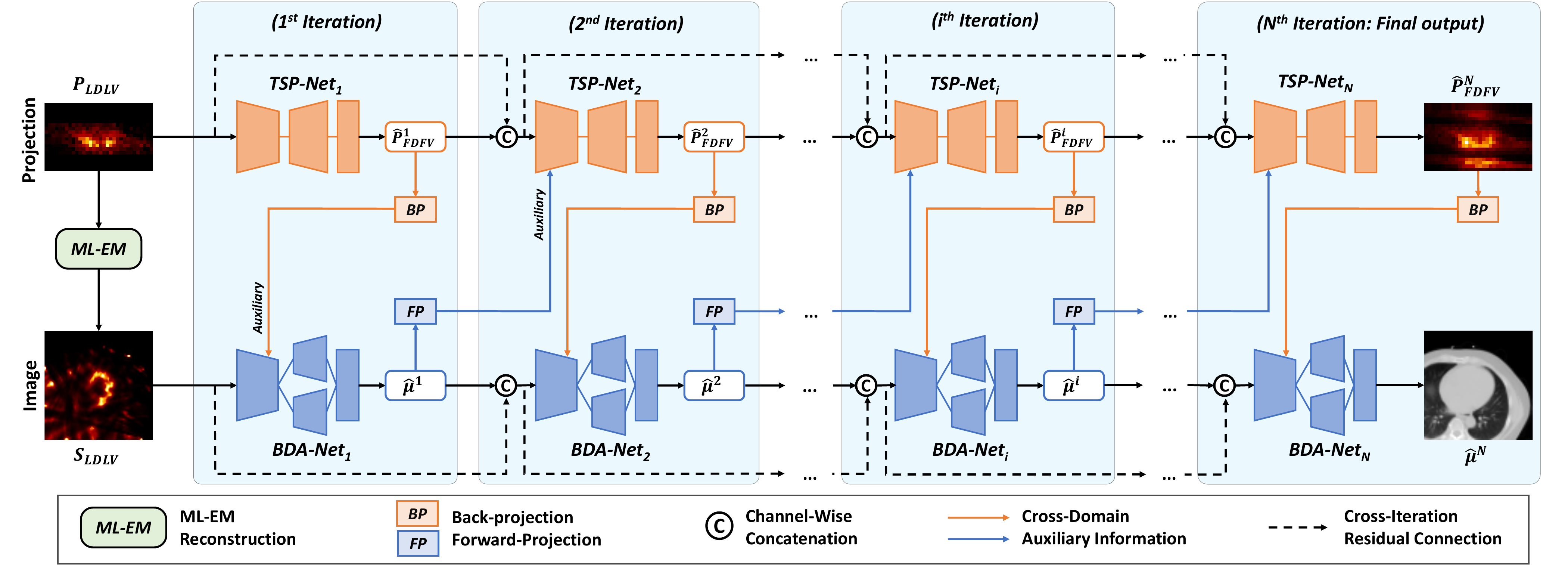}
\caption{Overview of the \textbf{Du}al-\textbf{Do}main \textbf{C}oarse-To-\textbf{F}ine Progressive \textbf{Net}work (DuDoCFNet). In each iteration, DuDoCFNet employs a Two-Stage Progressive Network (TSP-Net) in the projection domain for denoising and restoration of the LD and LV projections, and a Boundary-Aware Network (BDA-Net) in the image domain for predicting $\mu$-maps. All the TSP-Nets and BDA-Nets are cascaded to enable cross-domain and cross-modality feature fusion. The predicted projection and $\mu$-map of the last iteration are employed as the final prediction outputs of DuDoCFNet.}
\label{fig:overview}
\end{figure*}

Deep learning techniques have been developed to address either the low-dose (LD) denoising, LV reconstruction, or CT-free AC in nuclear medicine. Existing deep learning approaches for LD denoising in nuclear medicine are categorized into projection-domain and image-domain approaches. For the projection-domain approaches, Shiri \textit{et al.} \cite{shiri2020standard} applied a 2D ResNet to estimate full-dose (FD) projections from LD projections in cardiac SPECT. Similarly, Aghakhan \textit{et al.} \cite{aghakhan2022deep} used a 2D conditional GAN to denoise LD projections under multiple noise levels. Sun \textit{et al.} \cite{sun2022deep} further used a 3D conditional GAN to denoise LD projections in cardiac SPECT. In contrast, the image-domain approaches were developed by inputting LD images into neural networks to estimate FD images \cite{ramon2020improving, liu2021deep, liu2022personalized, yu2023need}. Previous studies proved that the projection-domain approaches consistently outperformed the image-domain approaches in both simulation and clinical studies \cite{sanaat2021deep, sun2022deep}.

Previous deep learning techniques for LV reconstruction in nuclear medicine can be summarized into projection-domain, image-domain, and dual-domain methods. In the projection-domain methods, Whiteley \textit{et al.} \cite{whiteley2019cnn} applied a U-Net \cite{ronneberger2015u} to estimate full-view (FV) projections from LV projections of whole-body Positron Emission Tomography (PET). Similarly, Shiri \textit{et al.} \cite{shiri2020standard} used a ResNet to predict FV projections from LV projections in cardiac SPECT. In contrast, the image-domain methods were developed by feeding reconstructed LV images into neural networks to estimate FV images \cite{amirrashedi2021deep}. Amirrashedi \textit{et al.} \cite{amirrashedi2021deep} further reported that the projection-domain methods outperformed the image-domain methods due to richer information in the projection presentation. Furthermore, Chen \textit{et al.} \cite{chen2023dudoss} proposed a dual-domain method named Dual-Domain Sinogram Synthesis (DuDoSS), which utilized the image-domain output as the prior information to estimate FV projection in the projection domain. DuDoSS outperformed both projection-domain and image-domain methods \cite{chen2023dudoss}. 

Deep learning-based CT-free AC in nuclear medicine can be generally classified into indirect and direct strategies \cite{chen2022deep}. For the indirect strategy, Shi \textit{et al.} \cite{shi2020deep} employed both U-Net and GAN to generate synthetic $\mu$-maps from SPECT images, which were then utilized for the AC reconstruction of cardiac SPECT. In contrast, Yang \textit{et al.} \cite{yang2021direct} applied a direct strategy in which the non-AC SPECT image was input to neural networks to estimate the AC SPECT image without the intermediate step of generating $\mu$-maps. Furthermore, the patient non-imaging clinical information was embedded to predict more accurate AC SPECT images in a recent study \cite{chen2022ct}. Previous works proved that the indirect approaches outperformed the direct approaches in multiple clinical scanners \cite{chen2022direct, du2022deep, chen2022cross}. However, the inaccurate estimation of the $\mu$-map boundaries remains a major limitation of the indirect approaches as shown in \cite{chen2022direct}.

\begin{figure*}[htb!]
\centering
\includegraphics[width=0.89\textwidth]{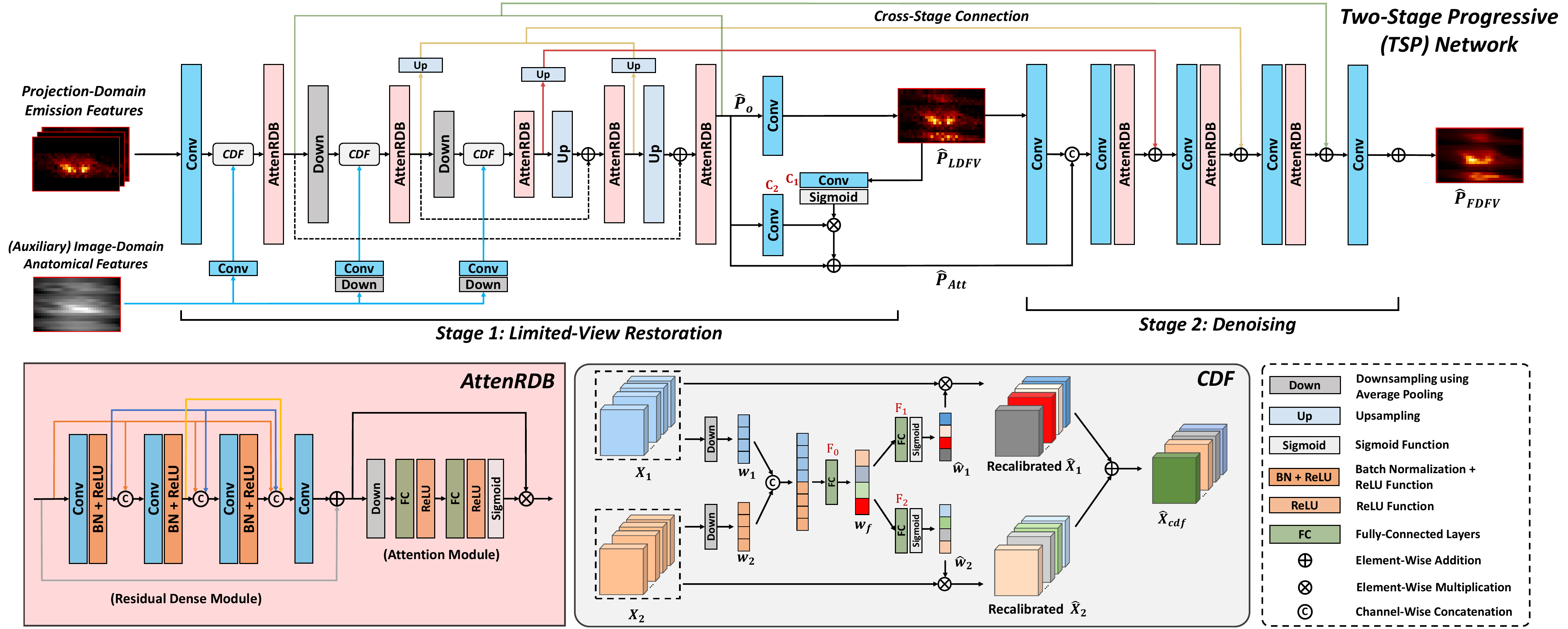}
\caption{Two-Stage Progressive Network (TSP-Net). In Stage 1, a U-Net-like structure is utilized to achieve the LV restoration. The auxiliary anatomical features are fed into multiple downsampling layers as the multi-layer fusion (MLF) mechanism. Cross-Domain Feature Fusion (CDF) modules recalibrate the channel weights for adaptive feature fusion. A non-downsampling module is employed in Stage 2 for the LD denoising.}
\label{fig:tsp}
\end{figure*}

While the aforementioned methods have been developed to individually tackle LD denoising, LV reconstruction, or CT-free AC in nuclear medicine, the solution for simultaneously addressing these tasks remains challenging and under-explored. Recently, multi-task simultaneous learning has been investigated in various medical imaging applications, which leverages domain-specific information across interrelated tasks to further enhance accuracy  \cite{zhou2021review}. For instance, existing strategies for simultaneous registration and segmentation involved utilizing a shared encoder for multi-task feature extraction \cite{qin2018joint, estienne2020deep} or employing segmentation results to assist the registration \cite{xu2019deepatlas, balakrishnan2019voxelmorph, ta2020semi, qiu2021rsegnet, li2021longitudinal}. Specifically, Qin \textit{et al.} \cite{qin2018joint} proposed a multi-scale network for the registration and segmentation of 2D cardiac Magnetic Resonance (MR) images. A shared encoder was applied by the segmentation and registration branches for feature extraction, leading to improved accuracy in both tasks. In contrast, Xu \textit{et al.} \cite{xu2019deepatlas} trained two convolutional modules for the registration and segmentation of 3D brain MR images, respectively. The registration module was supervised by an anatomy similarity loss based on the predicted masks from the segmentation module. Another important multi-task learning application is simultaneous segmentation and classification, where the common approach is utilizing segmentation results to improve the classification accuracy \cite{wu2018joint, li2022multir, hong2020mmcl, xu2020multi}. Specifically, Wu \textit{et al.} \cite{wu2018joint} applied a U-Net for the segmentation and classification of lung nodules in CT. The predicted segmentation masks and bottleneck features were fused to improve the subsequent classification. In addition, Xu \textit{et al.} \cite{xu2020multi} applied two cascaded convolutional modules, in which the output of the segmentation module was fed into the classification module to improve the classification of tongue images. The multi-task learning has also been explored in nuclear medicine imaging. Li \textit{et al.} \cite{li2021deep} proposed a recurrent framework for joint motion estimation and reconstruction in PET. In this study, a learned registration network was incorporated into a regularized PET image reconstruction module for simultaneous learning. Moreover, Zhou \textit{et al.} \cite{zhou2021mdpet} incorporated bidirectional LSTM layers into a Siamese pyramid network for simultaneous motion estimation and denoising in LD PET.

Although showing promising results, the above multi-task learning studies primarily focused on image-based features within single-domain frameworks. Dual-domain methods have exhibited superior performance to single-domain methods in various studies, due to augmented information constraints in both domains \cite{chen2023dudoss, chen2023dd, chen2022duald, chen2023dual}. Therefore, recent studies further explored fusing dual-domain features for better performance in multi-task learning \cite{zhou2022dudoufnet, zhou2022dudodr}. However, these studies were conducted based on a single imaging modality. Fusing multi-modality image features was proved to be more effective than using a single modality in deep learning applications \cite{wang2021review} since the complementary information from multiple modalities enables more comprehensive feature representations. Multi-modality information fusion have been explored in a wide range of deep learning studies based on CNN \cite{liu2017medical, hou2019brain, liu2019medical, xia2019novel, song2022cross, chen2023dusfe}. Liu et al. \cite{liu2017medical} presented a Siamese convolutional network to obtain a weighted map that fuses the pixel-wise information of the two input images. Hou et al. \cite{hou2019brain} introduced a multi-modality fusion method based on a dual-channel spiking cortical model (DCSCM), which generated and combined both low- and high-frequency coefficients of images to achieve enhanced feature fusion. Xia et al. \cite{xia2019novel} proposed a fusion scheme for multi-modality medical images that utilizes and combines the image features from both the multi-scale transformation and convolutional modules. Song et al. \cite{song2022cross} presented a cross-attention block based on non-local attention for the cross-modality fusion and registration of ultrasound and CT images. Thus, a recent study proposed a cross-domain and cross-modality network (CDI-Net) for multi-task learning \cite{chen2023cross}. However, the simple channel concatenation in CDI-Net might not effectively fuse the cross-domain or cross-modality image features. Furthermore, the basic U-Net models within the CDI-Net might not be optimal for estimating projections or $\mu$-maps, without considering the characteristics of specific images and tasks.

Therefore, we propose a \textbf{Du}al-\textbf{Do}main \textbf{C}oarse-to-\textbf{F}ine Progressive \textbf{Net}work (DuDoCFNet) for simultaneous LD denoising, LV reconstruction, and $\mu$-map generation of cardiac SPECT. Paired projection-domain and image-domain networks are cascaded using a multi-layer fusion (MLF) mechanism for cross-domain and cross-modality feature fusion. In the projection domain, Two-Stage Progressive Networks (TSP-Net) are utilized for LD denoising and LV restoration. The U-Net-like downsampling-upsampling framework \cite{ronneberger2015u} can restore general structures but might not preserve finer image details. Thus, TSP-Net utilizes a U-Net-like framework in Stage 1 to restore the coarse LV structures, followed by a non-downsampling module in Stage 2 to recover finer details of the LD projection. In the image domain, Boundary-Aware Networks (BDA-Net) enhance spatial attention on image boundaries and thus improve the boundary accuracy of the predicted $\mu$-maps. BDA-Net first employs a shared encoder in Stage 1 to predict a coarse $\mu$-map and its boundary image, which are then adaptively fused in Stage 2 to generate a refined $\mu$-map. DuDoCFNet was trained end-to-end using both projection and image losses. Experiments showed that DuDoCFNet achieved superior accuracy under various iterations and LD levels.

\begin{figure*}[htb!]
\centering
\includegraphics[width=0.91\textwidth]{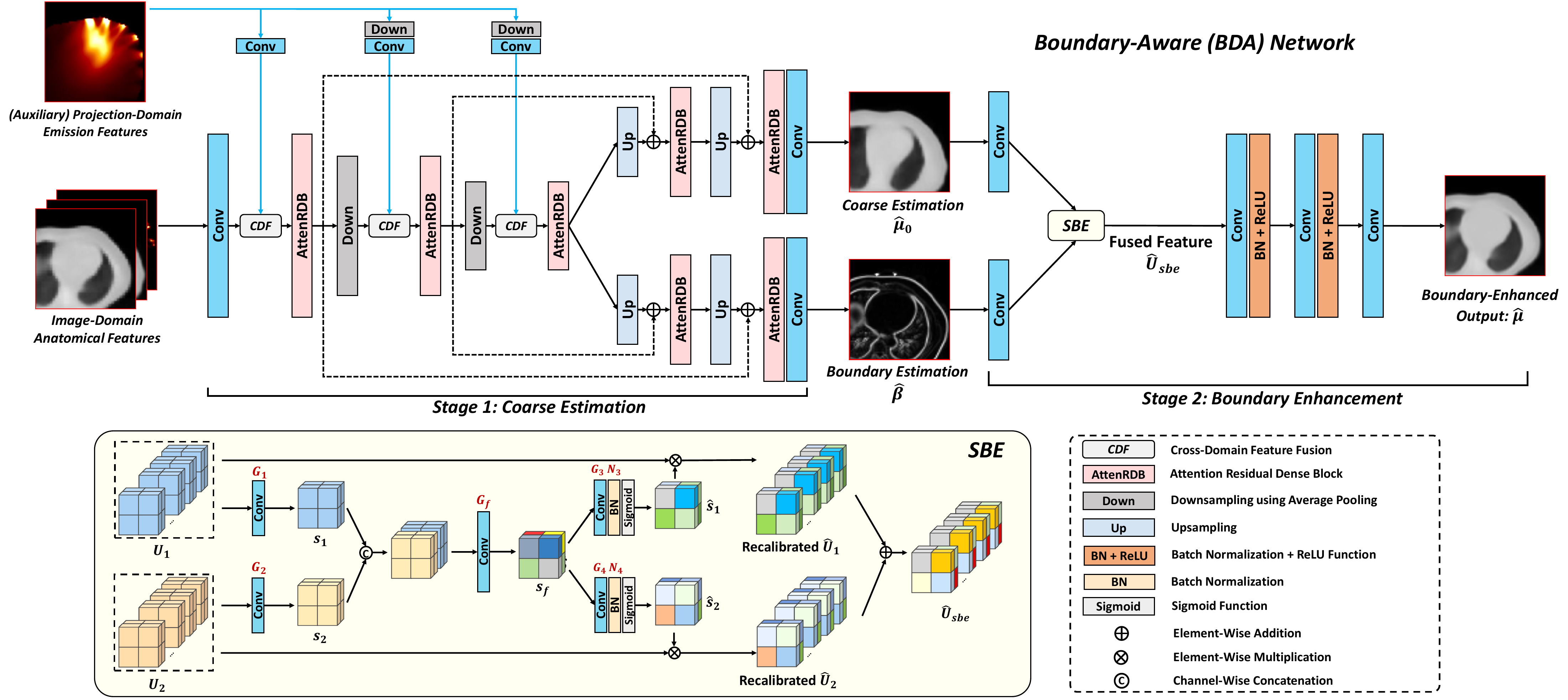}
\caption{Boundary-Aware Network (BDA-Net). A shared encoder and two task-specific decoders are utilized to estimate a coarse $\mu$-map and its boundary image. Cross-domain features are embedded in multiple downsampling layers as the multi-level fusion. The estimated $\mu$-map and boundary image are jointly fed into a Spatial Boundary Enhancement (SBE) module to enhance the boundary accuracy of the final refined $\mu$-map.}
\label{fig:bda}
\end{figure*}

\section{Methods}
\subsection{Problem Formulation}
The goal of this multi-task learning study is to generate the predicted FD and FV projection ($\hat{P}_{FDFV}$) and $\mu$-map ($\hat{\mu}$) with the LD and LV projection ($P_{LDLV}$) as input, formulated as:
\begin{equation}
\label{eq:1}
   [ \hat{P}_{FDFV}, \hat{\mu} ] = \mathcal{H} \left( P_{LDLV}  \right), 
\end{equation}
where $\mathcal{H}\left( \cdot \right)$ is the DuDoCFNet operator. The output labels are the ground-truth FD and FV projection ($P_{FDFV}$) and the CT-derived $\mu$-map ($\mu$). Then, $\hat{P}_{FDFV}$ and $\hat{\mu}$ are input into a reconstruction module to output the predicted AC SPECT images. Thus, predicting $\hat{P}_{FDFV}$ achieves the LD denoising and LV restoration, and predicting $\hat{\mu}$ enables the CT-free AC.

\subsection{Data Preparation}
A total of 600 anonymized clinical hybrid one-day SPECT-CT stress/rest MPI studies were included in this work. Each study was acquired on a GE NM/CT 570c dedicated SPECT-CT scanner following the injection of $^{99\mathrm{m}}$Tc-tetrofosmin. The GE 570c system has 19 pinhole detectors arranged in three columns on a cylindrical surface. The top, center, and bottom columns contain 5, 9, and 5 detectors, respectively \cite{chan2016impact}. 

The LV projection was produced by including the central 9-angle projection and zero-padding the peripheral angles, simulating the latest GE MyoSPECT ES system \cite{gehealthcaremyospectessystems}. The LD projection was generated by randomly decimating the FD list-mode data with a default 10$\%$ downsampling rate. $P_{LDLV} \in \mathbb{R}^{32\times32\times19}$ was produced by conducting both the LV and LD downsampling. $P_{FDFV} \in \mathbb{R}^{32\times32\times19}$ was the original FD and FV projection. The CT-derived $\mu$-maps ($\mu\in\mathbb{R}^{72\times72\times40}$) with a voxel size of $4\times4\times4~\mathrm{mm^3}$ were well registered and resolution-matched to the SPECT images. 250, 100, and 250 cases were utilized for training, validation, and testing.

\subsection{DuDoCFNet Overview}
The framework of DuDoCFNet is presented in Fig. \ref{fig:overview}. $P_{LDLV}$ is first input to a Maximum-Likelihood Expectation-Maximization module (ML-EM, 30 iterations) to reconstruct the LD and LV SPECT image $S_{LDLV}$. Then, $P_{LDLV}$ and $S_{LDLV}$ are fed into DuDoCFNet to simultaneously estimate $P_{FDFV}$ and $\mu$. The projection-domain \textit{TSP-Nets} (Fig. \ref{fig:tsp}) and image-domain \textit{BDA-Nets} (Fig. \ref{fig:bda}) are cascaded for cross-domain and cross-modality feature fusion. The details of \textit{TSP-Net} and \textit{BDA-Net} are described in subsequent subsections.

In the $1^{st}$ iteration, $P_{LDLV}$ is first input into \textit{TSP-Net$_1$} and output the predicted FD and FV projection $\hat{P}^{1}_{FDFV}$. After Back-Projection (BP), $\hat{P}^{1}_{FDFV}$ is input to \textit{BDA-Net$_1$}, leveraging the auxiliary emission information to enhance the $\mu$-map estimation. $S_{LDLV}$ is also input into \textit{BDA-Net$_1$} to provide the image-domain information. An MLF mechanism (to be described later) is employed in \textit{BDA-Net$_1$} to adaptively fuse the cross-domain features to produce a more accurate $\hat{\mu}^{1}$.

\begin{figure*}[htb!]
\centering
\includegraphics[width=0.92\textwidth]{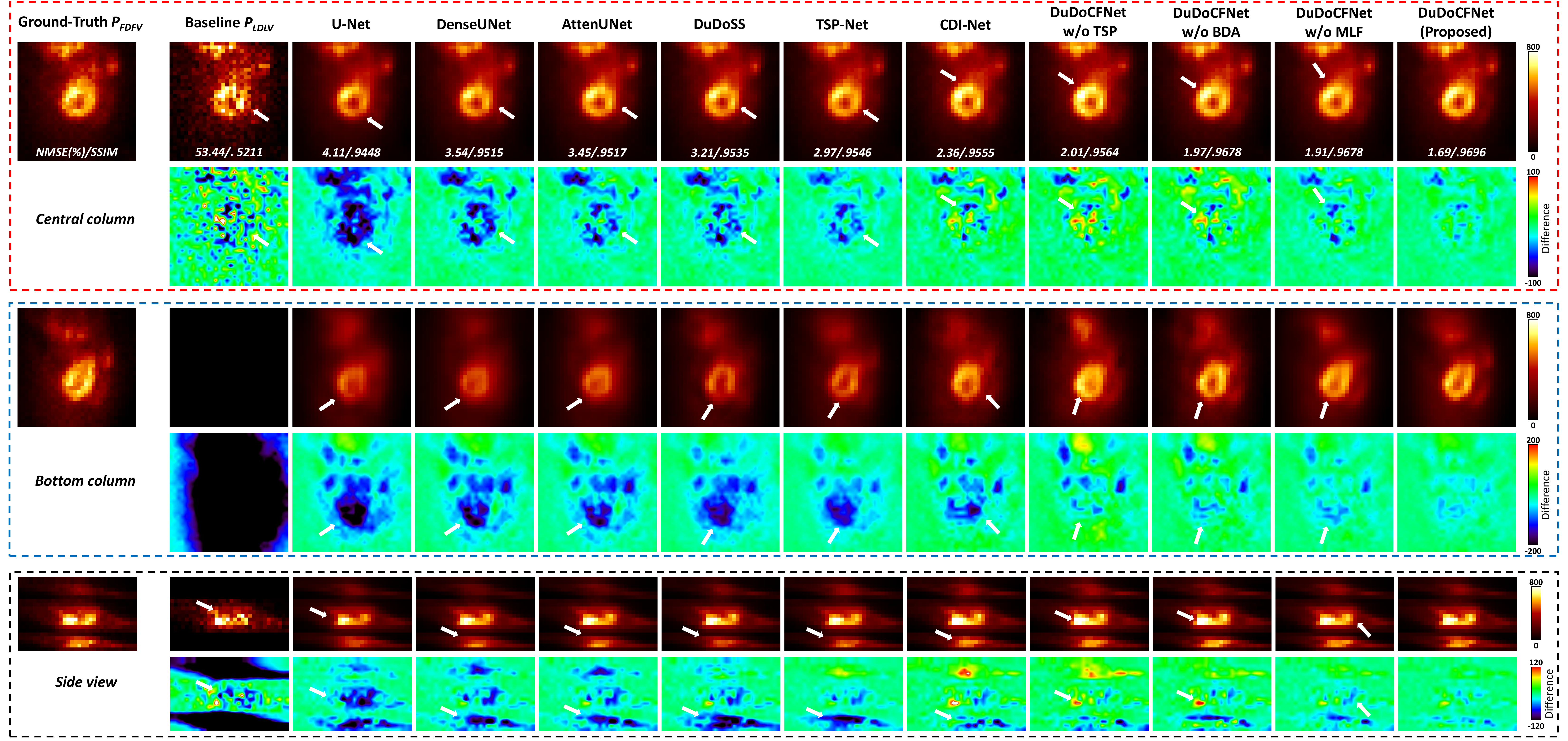}
\caption{Predicted FD and FV projections displayed in the central-column angle, bottom-column angle, and side view. White arrows denote the regions with over- or under-estimated projection intensities. NMSE and SSIM between the predicted and ground-truth projections are annotated.}
\label{fig:proj}
\end{figure*}

Then, in the $i^{th} (i\geq2)$ iteration, the image-domain output of the previous iteration, $\hat{\mu}^{(i-1)}$, is input to \textit{TSP-Net$_i$} after Forward-Projection (FP). This employs the auxiliary anatomical information of $\hat{\mu}^{(i-1)}$ to assist the projection estimation. Additionally, the predicted projections from previous ($i-1$) iterations are concatenated and input into \textit{TSP-Net$_i$}, preserving the previously estimated emission features to gradually enhance the prediction accuracy. \textit{TSP-Net$_i$} also utilizes the MLF mechanism to adaptively fuse the cross-domain and cross-modality features. The output of \textit{TSP-Net$_i$} is formulated as:
\begin{equation}
\label{eq:2}
\resizebox{0.91\hsize}{!}{$
    \hat{P}^{i}_{FDFV} = \mathcal{T}_i\left( \Gamma_f\left(\hat{\mu}^{(i-1)}\right), \, \left\{\hat{P}^{1}_{FDFV}, \, \cdots, \, \hat{P}^{(i-1)}_{FDFV}, \; P_{LDLV}\right\}  \right),
$}
\end{equation}
where $\mathcal{T}_i$ is the \textit{TSP-Net$_i$} and $\Gamma_f$ is the FP operator. $\left\{ \cdot \right\}$ refers to channel-wise concatenation. Similarly, $\hat{P}^{i}_{FDFV}$ after BP is input to \textit{BDA-Net$_i$}, employing the auxiliary emission information of $\hat{P}^{i}_{FDFV}$ to assist the $\mu$-map estimation. The predicted $\mu$-maps from previous ($i-1$) iterations are also concatenated and input into \textit{BDA-Net$_i$}, generating $\hat{\mu}^i$ as:
\begin{equation}
\label{eq:3}
\resizebox{0.85\hsize}{!}{$
    \hat{\mu}^i = \mathcal{D}_i\left(\Gamma_b(\hat{P}^{i}_{FDFV}), \; \left\{\hat{\mu}^1, \, \cdots, \, \hat{\mu}^{(i-1)}, \, S_{LDLV} \right\} \right),
$}
\end{equation}
where $\mathcal{D}_i$ is the \textit{BDA-Net$_i$} and $\Gamma_b$ is the BP operator. The predictions of the $N^{th}$ iteration ($N$ represents the total number of iterations, default 5), $\hat{P}^{N}_{FDFV}$ and $\hat{\mu}^N$, are the prediction outputs of DuDoCFNet as described in Eq. \ref{eq:1}.

\subsection{TSP-Net in Projection Domain}
The architecture of TSP-Net is illustrated in Fig. \ref{fig:tsp}. As mentioned above, the downsampling-upsampling framework can recover general structures but not finer image details. Thus, TSP-Net uses a U-Net-like framework for coarse LV restoration in Stage 1 and a non-downsampling module for finer LD denoising in Stage 2.  As described in Eq. \ref{eq:2}, the inputs to TSP-Net include the projection-domain emission features and the auxiliary image-domain anatomical features after FP.

In Stage 1, residual dense blocks with attention (AttenRDB, depicted at the bottom left of Fig. \ref{fig:tsp}) are employed for image feature extractions in TSP-Net. The emission and the auxiliary anatomical features are connected at multiple downsampling layers to enable effective feature fusion in various spatial dimensions, which is the MLF mechanism. Due to the discrepancy between the two imaging modalities, we propose a Cross-Domain Feature Fusion (CDF) module to calibrate the channel-wise weights before fusing the two-modality features. As shown at the bottom right of Fig. \ref{fig:tsp}, the two inputs of CDF, $X_1$ and $X_2 \in \mathbb{R}^{h \times w \times d \times c}$ ($h$, $w$, $d$, and $c$ refer to height, width, depth, and number of channels), are first downsampled into $w_1$ and $w_2 \in \mathbb{R}^{c}$ to encode the channel-wise weight features. Then, $w_1$ and $w_2$ are fused into $w_f \in \mathbb{R}^{2c}$ using concatenation and a fully-connected layer, described as:
\begin{equation}
\resizebox{0.45\hsize}{!}{$
    w_f = F_0 \left( \left\{ D(X_1), \;  D(X_2) \right\} \right),
$}
\end{equation}
where $D(\cdot)$ is downsampling by average pooling and $F_0$ is the fully-connected layer. Then, $w_f$ is utilized to generate the recalibration weights $\hat{w}_1$ and $\hat{w}_2\in \mathbb{R}^{c}$, which are applied back to $X_1$ and $X_2$ using element-wise multiplication. The recalibrated features are then concatenated to generate the CDF output $\hat{X}_{cdf} \in \mathbb{R}^{h \times w \times d \times 2c}$, formulated as:
\begin{equation}
\resizebox{0.68\hsize}{!}{$
    \hat{X}_{cdf} = \left\{ \sigma(F_1 (w_f)) \otimes X_1, \; \sigma(F_2 (w_f)) \otimes X_2    \right\},
$}
\end{equation}
where $\sigma(\cdot)$ is the sigmoid activation function and $\otimes$ refers to the element-wise multiplication. Therefore, CDF effectively fuses the channel information and optimizes the fusion weights of the anatomical and emission features. Next, the output of the last decoding layer, $\hat{P}_0$, goes through a $1\times1\times1$ convolution layer to generate the predicted LD and FV projection $\hat{P}_{LDFV}$.

In Stage 2, $\hat{P}_{LDFV}$ is input to a non-downsampling module for denoising. Meanwhile, we apply a self-attention block to refine and improve the $\hat{P}_{LDFV}$ before feeding it to the next stage, inspired by \cite{zamir2021multi}. The self-attention output is:
\begin{equation}
\resizebox{0.60\hsize}{!}{$
    \hat{P}_{Att} = \hat{P}_0 + C_2(\hat{P}_0) \otimes \sigma(C_1(\hat{P}_{LDFV})),
$}
\end{equation}
where $C_1(\cdot)$ and $C_2(\cdot)$ are $1\times1\times1$ convolution. $\hat{P}_{Att}$ is fed to Stage 2 for feature reserving. In addition, as depicted at the top of Fig. \ref{fig:tsp}, the image features from multiple layers of Stage 1 are concatenated and input into consecutive AttenRDB layers in Stage 2. This utilizes the information from Stage 1 in multiple spatial dimensions to gradually enhance prediction accuracy in Stage 2. Finally, $\hat{P}_{FDFV}$ is generated as the output of Stage 2. The training loss of \textit{TSP-Net$_i$} is calculated based on its outputs from both stages, formulated as:
\begin{equation}
\label{eq:7}
\resizebox{0.85\hsize}{!}{$
    \mathcal{L}^i_{proj} = \Vert \hat{P}^i_{LDFV} - P_{LDFV} \Vert_1 + \Vert \hat{P}^i_{FDFV} - P_{FDFV} \Vert_1,
$}
\end{equation}
where $\hat{P}^i_{LDFV}$ and $\hat{P}^i_{FDFV}$ indicate the predictions in the $i^{th}$ iteration. $P_{LDFV}$ is the ground-truth LD and FV projection.

\begin{table} [htb!]
\caption{Quantitative evaluations of predicted projections on 250 testing cases using NMSE, SSIM, and PSNR. The numbers of network parameters (Param, unit: million) are indicated in the last column. The best results are marked in \textbf{BOLD}.}
\label{tab:proj} 
\footnotesize
\centering
\resizebox{0.47\textwidth}{!}{
\begin{tabular}{ l | c | c | c || c || c}
\hline
\textbf{Methods}  & \textbf{NMSE ($\boldsymbol{\%}$)}    & \textbf{SSIM}   & \textbf{PSNR}  & \textbf{P-values$^{a}$}  & \textbf{Param (m)} \Tstrut\Bstrut\\  
\hline  
Baseline LDLV                   & $54.35 \pm 2.34$     & $.492 \pm .025$        & $19.14 \pm 1.69$       & $<$ 0.001      & \textendash  \Tstrut\Bstrut\\
\hline 
U-Net \cite{ronneberger2015u}        & $4.21 \pm 1.41$      & $.923 \pm .024$        & $30.45 \pm 1.46$       & $<$ 0.001      & 1.36  \Tstrut\Bstrut\\  
\hline
DenseUNet \cite{huang2017densely}   & $3.26 \pm 1.02$      & $.938 \pm .015$        & $31.55 \pm 1.50$       & $<$ 0.001      & 1.43  \Tstrut\Bstrut\\  
\hline
AttenUNet \cite{oktay2018attention} & $3.11 \pm 1.01$      & $.943 \pm .015$        & $31.77 \pm 1.45$       & $<$ 0.001      & 1.54  \Tstrut\Bstrut\\  
\hline
DuDoSS \cite{chen2023dudoss}        & $2.92 \pm 0.96$      & $.943 \pm .016$        & $32.04 \pm 1.49$       & $<$ 0.001      & 3.09  \Tstrut\Bstrut\\  
\hline 
TSP-Net [Fig. \ref{fig:tsp}]  & $2.66 \pm 0.81$      & $.949 \pm .015$        & $32.43 \pm 1.65$       & $<$ 0.001      & 1.01  \Tstrut\Bstrut\\  
\hline
CDI-Net \cite{chen2023cross}        & $2.28 \pm 0.74$      & $.950 \pm .015$        & $33.12 \pm 1.66$       & $<$ 0.001      & 3.11$\times$N$^{b}$  \Tstrut\Bstrut\\  
\hline \hline
DuDoCFNet w/o TSP                 & $2.15 \pm 0.64$      & $.955 \pm .013$        & $33.34 \pm 1.73$       & $<$ 0.001      & 1.42$\times$N  \Tstrut\Bstrut\\  
\hline
DuDoCFNet w/o BDA                 & $2.12 \pm 0.65$      & $.954 \pm .013$        & $33.42 \pm 1.68$       & $<$ 0.001      & 1.52$\times$N  \Tstrut\Bstrut\\  
\hline
DuDoCFNet w/o MLF                 & $2.09 \pm 0.68$      & $.955 \pm .013$        & $33.51 \pm 1.68$       & $<$ 0.001      & 1.76$\times$N  \Tstrut\Bstrut\\  
\hline
\textbf{DuDoCFNet (Proposed)}           & $\mathbf{1.83 \pm 0.57}$  & $\mathbf{.958 \pm .013}$    & $\mathbf{34.07 \pm 1.68}$  & \textendash  & 1.77$\times$N    \Tstrut\Bstrut\\  
\hline
\multicolumn{6}{l}{$^{a}$P-values of paired t-tests on NMSE between the current testing group and DuDoCFNet (Proposed).} \\
\multicolumn{6}{l}{$^{b}$Number of iterations of the deep learning framework with a default of 5.} \\
\end{tabular}
}
\end{table}

\begin{figure*}[htb!]
\centering
\includegraphics[width=0.92\textwidth]{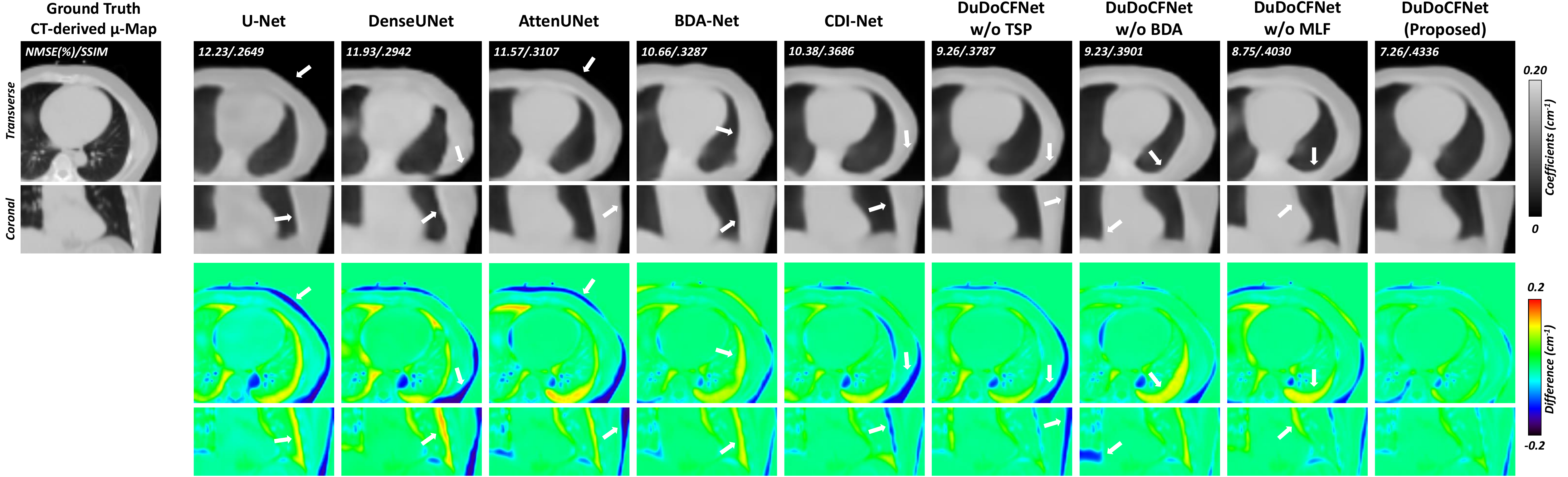}
\caption{Predicted $\mu$-maps (unit: cm$^{-1}$) with error maps. White arrows denote the $\mu$-map regions with inaccurate estimations. DuDoCFNet demonstrates the most accurate boundary estimations. NMSE and SSIM between the predicted and ground-truth $\mu$-maps are annotated.}
\label{fig:amap}
\end{figure*}

\subsection{BDA-Net in Image Domain}
The structure of BDA-Net is shown in Fig. \ref{fig:bda}. As mentioned above, the inaccurate boundary estimation poses a significant limitation in predicting $\mu$-maps \cite{chen2022direct}. To address this, BDA-Net first uses a shared encoder in Stage 1 to predict a coarse $\mu$-map and a boundary image as the preliminary results. Then, the predicted coarse $\mu$-map and boundary image are fused in Stage 2 to enhance the spatial attention on the image boundaries, thus generating a refined $\mu$-map. As described in Eq. \ref{eq:3}, the inputs to BDA-Net include the image-domain anatomical features and the auxiliary projection-domain emission features after BP.

In the shared encoder of Stage 1, the MLF mechanism is utilized to connect and fuse the anatomical and the auxiliary emission features at multiple downsampling layers. Then, inspired by \cite{estienne2020deep}, two task-specific decoders are used to predict a coarse image $\hat{\mu}_0$ and a boundary profile $\hat{\beta}$. Next, a Spatial Boundary Enhancement (SBE) module utilizes $\hat{\mu}_0$ and $\hat{\beta}$ as inputs to fuse the spatial features and enhance the spatial attention on the image boundaries. As shown at the bottom of Fig. \ref{fig:bda}, the input $U_1$ and $U_2 \in \mathbb{R}^{72 \times 72 \times 40 \times c}$ are squeezed into $s_1$ and $s_2 \in \mathbb{R}^{72 \times 72 \times 40 \times 1}$ using $3\times3\times3$ convolutions. Then, $s_1$ and $s_2$ are fused by concatenation and a convolutional layer, generating the combined spatial feature $s_f \in \mathbb{R}^{72 \times 72 \times 40 \times 2}$ as:
\begin{equation}
\resizebox{0.45\hsize}{!}{$
    s_f = G_f(\left\{ G_1(U_1), \; G_2(U_2) \right\}),
$}
\end{equation}
where $G_1(\cdot)$, $G_2(\cdot)$, $G_f(\cdot)$ refer to the $3\times3\times3$ convolutions. Then, $\hat{s}_1$ and $\hat{s}_2$ are generated from $s_f$ and applied back to $U_1$ and $U_2$ using element-wise multiplication. Then, the recalibrated spatial features are concatenated to generate the SBE output $\hat{U}_{sbe} \in \mathbb{R}^{72 \times 72 \times 40 \times 2c}$, formulated as:
\begin{equation}
\resizebox{0.80\hsize}{!}{$
    \hat{U}_{sbe} = \left\{ \sigma(N_3(G_3(s_f))) \otimes U_1, \;  \sigma(N_4(G_4(s_f))) \otimes U_2          \right\},
$}
\end{equation}
where $G_3(\cdot)$ and $G_4(\cdot)$ refer to convolutions. $N_3(\cdot)$ and $N_4(\cdot)$ are the batch normalization layers. Then, $\hat{U}_{sbe}$ is input to the subsequent convolutional module in Stage 2 to extract the boundary information. Finally, a refined $\mu$-map with more accurate boundaries, $\hat{\mu}$, is generated as the output of BDA-Net. The training loss of \textit{BDA-Net$_i$} is calculated based on its preliminary and refined outputs in both stages, formulated as:
\begin{equation}
\label{eq:10}
    \mathcal{L}^i_{img} = \Vert \hat{\beta}^i - \beta \Vert_1  +  \Vert \hat{\mu}^i_0 - \mu \Vert_1  +  \Vert \hat{\mu}^i - \mu \Vert_1,
\end{equation}
where $\hat{\beta}^i$, $\hat{\mu}^i_0$, and $\hat{\mu}^i$ are predicted boundary and $\mu$-maps in the $i^{th}$ iteration. $\beta$ represents the boundary of $\mu$ and is determined by calculating the 3D gradients of $\mu$, formulated as:
\begin{equation}
    \beta = \Vert \nabla \mu_x \Vert_2 + \Vert \nabla \mu_y \Vert_2 + \Vert \nabla \mu_z \Vert_2.
\end{equation}

\begin{table} [htb!]
\caption{Quantitative evaluations of predicted $\mu$-maps on 250 testing cases using NMSE, SSIM, and PSNR. The numbers of network parameters (Param, unit: million) are indicated in the last column. The best results are marked in \textbf{BOLD}.}
\label{tab:amap} 
\footnotesize
\centering
\resizebox{0.47\textwidth}{!}{
\begin{tabular}{ l | c | c | c || c || c}
\hline
\textbf{Methods}  & \textbf{NMSE ($\boldsymbol{\%}$)}   & \textbf{SSIM}   & \textbf{PSNR}  & \textbf{P-values$^{a}$}  & \textbf{Param (m)} \Tstrut\Bstrut\\  
\hline  
U-Net \cite{ronneberger2015u}        & $12.96 \pm 4.90$      & $.237 \pm .047$        & $17.26 \pm 1.82$       & $<$ 0.001      & 1.36  \Tstrut\Bstrut\\  
\hline
DenseUNet \cite{huang2017densely}   & $12.75 \pm 5.13$      & $.242 \pm .048$        & $17.30 \pm 1.83$       & $<$ 0.001      & 1.43  \Tstrut\Bstrut\\  
\hline
AttenUNet \cite{oktay2018attention} & $12.51 \pm 4.33$      & $.247 \pm .050$        & $17.33 \pm 1.82$       & $<$ 0.001      & 1.54  \Tstrut\Bstrut\\  
\hline
BDA-Net [Fig. \ref{fig:bda}]    & $12.34 \pm 4.46$      & $.253 \pm .048$        & $17.41 \pm 1.84$       & $<$ 0.001      & 0.77  \Tstrut\Bstrut\\  
\hline
CDI-Net \cite{chen2023cross}        & $12.19 \pm 5.04$      & $.258 \pm .047$        & $17.51 \pm 1.89$       & $<$ 0.001      & 3.11$\times$N$^{b}$  \Tstrut\Bstrut\\  
\hline \hline
DuDoCFNet w/o TSP                 & $11.91 \pm 4.77$      & $.264 \pm .057$        & $17.60 \pm 1.84$       & $<$ 0.001      & 1.42$\times$N  \Tstrut\Bstrut\\  
\hline
DuDoCFNet w/o BDA                 & $11.97 \pm 4.90$      & $.263 \pm .057$        & $17.58 \pm 1.91$       & $<$ 0.001      & 1.52$\times$N  \Tstrut\Bstrut\\  
\hline
DuDoCFNet w/o MLF                 & $11.89 \pm 4.72$      & $.265 \pm .055$        & $17.60 \pm 1.88$       & $<$ 0.001      & 1.76$\times$N  \Tstrut\Bstrut\\  
\hline
\textbf{DuDoCFNet (Proposed) }          & $\mathbf{11.43 \pm 4.67}$   & $\mathbf{.270 \pm .058}$    & $\mathbf{17.83 \pm 1.85}$  & \textendash  & 1.77$\times$N    \Tstrut\Bstrut\\  
\hline
\multicolumn{6}{l}{$^{a}$P-values of paired t-tests on NMSE between the current testing group and DuDoCFNet (Proposed).} \\
\multicolumn{6}{l}{$^{b}$Number of iterations of the deep learning framework with a default of 5.} \\
\end{tabular}
}
\end{table}

\begin{figure*}[htb!]
\centering
\includegraphics[width=0.90\textwidth]{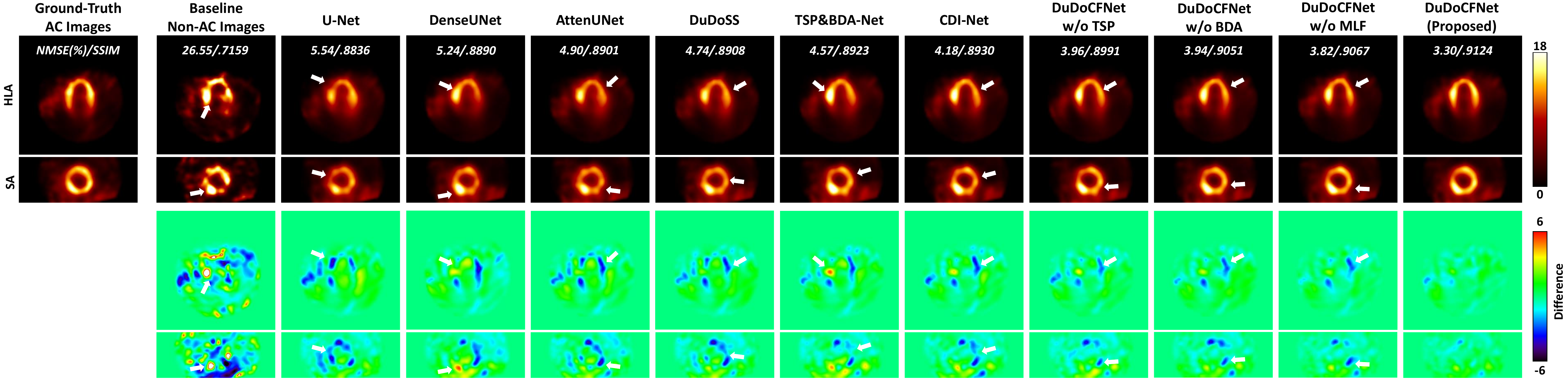}
\caption{Reconstructed AC SPECT images using predicted projections and $\mu$-maps. White arrows denote the image regions with inaccurate reconstructions. DuDoCFNet outputs the most accurate AC images. NMSE and SSIM between predicted and ground-truth images are annotated.}
\label{fig:ac}
\end{figure*}

\begin{figure*}[htb!]
\centering
\includegraphics[width=0.85\textwidth]{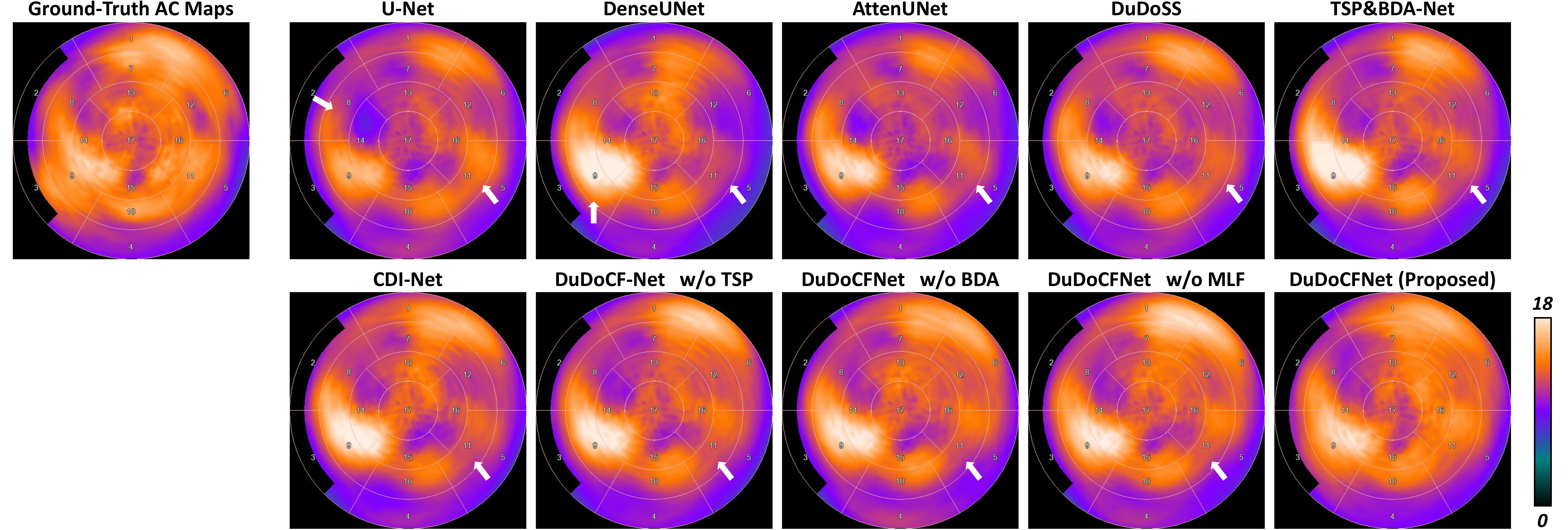}
\caption{Standard 17-segment polar maps of the AC SPECT images. White arrows denote the segment regions with over- or under-estimated intensities. The polar map by DuDoCFNet is the most consistent with ground truth, compared to single-task, CDI-Net, and ablation study groups.}
\label{fig:ac_polar}
\end{figure*}

\subsection{Implementation Details and Ablation Studies}
DuDoCFNet is end-to-end trained using both projection and image losses in all iterations. The overall loss function is:
\begin{equation}
    \mathcal{L} = \sum_{i=1}^{N}( \alpha_{p} \, \mathcal{L}^i_{proj} + \alpha_{i} \, \mathcal{L}^i_{img} ),
\end{equation}
where N is the total number of iterations (default: 5). $\mathcal{L}^i_{proj}$ and $\mathcal{L}^i_{img}$ are the projection and image losses described in Eq. \ref{eq:7} and \ref{eq:10}. $\alpha_{p}$ and $\alpha_{i}$ are loss weights. DuDoCFNet achieves the optimal performance when $\alpha_{p}=1$ and $\alpha_{i}=0.2$.

In this study, DuDoCFNet was compared to various single- and multi-task learning methods. For estimating projections, U-Net \cite{ronneberger2015u}, Densely-Connected U-Net (DenseUNet) \cite{huang2017densely}, and Attention U-Net (AttenUNet) \cite{oktay2018attention} were employed to estimate $P_{FDFV}$ using $P_{LDLV}$ as the input. DuDoSS \cite{chen2023dudoss} was included as the dual-domain method. For the $\mu$-map estimation, U-Net, DenseUNet, and AttenUNet were utilized to predict $\mu$ using $S_{LDLV}$ as the input. Moreover, the multi-task learning method, CDI-Net \cite{chen2023cross}, was applied to simultaneously estimate $P_{FDFV}$ and $\mu$. CDI-Net is set to run for 5 iterations as in \cite{chen2023cross}.

Ablation studies of DuDoCFNet were also conducted for further evaluation. Either TSP-Net or BDA-Net was solely used to predict $P_{FDFV}$ or $\mu$, to assess the impact of the multi-task learning framework on the prediction accuracy. Furthermore, to assess the effect of the progressive learning strategy, Stage 2 of TSP-Net or BDA-Net in DuDoCFNet is removed to produce DuDoCFNet w/o TSP and DuDoCFNet w/o BDA. Moreover, the MLF mechanism is replaced by input-channel concatenations as in \cite{chen2023cross} to produce DuDoCFNet w/o MLF. All networks were implemented using PyTorch \cite{paszke2019pytorch} with Adam optimizers. The projection-domain and image-domain modules were trained with learning rates of $10^{-3}$ and $10^{-4}$ and a batch size of 2. The single-task learning networks were trained for 200 epochs, while CDI-Net and DuDoCFNet were only trained for 50 epochs to reach full convergence. To minimize the computational costs of the iterative network, our DuDoCFNet is designed to be super lightweight by reducing the number of channels in its convolutional layers.

\section{Results}
\subsection{Predicted Projections}  
Normalized Mean Square Error (NMSE), Structural Similarity (SSIM), and Peak Signal-to-Noise Ratio (PSNR) are used for voxel-wise quantitative evaluations of predicted projections, $\mu$-maps, and reconstructed AC images. Clinical 17-segment polar maps are generated from AC images and quantified using Absolute Percent Error (APE) \cite{chen2022ct}.

Fig. \ref{fig:proj} shows the predicted FD and FV projections. We can observe that the single-task learning methods including U-Net, DenseUNet, AttenUNet, and DuDoSS significantly underestimate the projection intensities in the cardiac regions. In contrast, TSP-Net outputs more accurate projections than AttenUNet, showing the effectiveness of the progressive learning framework (Fig. \ref{fig:tsp}) in recovering the FD and FV projections. The multi-task learning method, CDI-Net, produces more accurate projections than single-task learning methods. Moreover, DuDoCFNet outperforms CDI-Net and ablation groups, confirming the efficacy of our proposed progressive learning strategies and the MLF mechanism in enhancing estimation accuracy. Table \ref{tab:proj} lists the quantitative evaluations of the predicted projections. It can be observed that TSP-Net shows higher accuracy than other single-task learning methods but uses fewer parameters. DuDoCFNet exhibits higher accuracy than ablation study groups and CDI-Net (NMSE, 1.83$\%$ versus 2.28$\%$, p$<$0.001) but has fewer parameters than CDI-Net.

\subsection{Predicted Attenuation Maps}  
Fig. \ref{fig:amap} shows the predicted $\mu$-maps. The indirect approaches using U-Net, DenseUNet, and AttenUNet exhibit inaccurate estimations of $\mu$-map boundaries. BDA-Net estimates a more accurate $\mu$-map than AttenUNet, proving that the boundary enhancement mechanism (Fig. \ref{fig:bda}) improves the prediction accuracy. The predicted $\mu$-map by DuDoCFNet is more consistent with ground truth compared to CDI-Net and the ablation groups. This demonstrates that the progressive learning strategies and the MLF mechanism in DuDoCFNet improve prediction accuracy. Table \ref{tab:amap} illustrates the quantitative evaluations of the predicted $\mu$-maps. BDA-Net generates more accurate $\mu$-maps than other indirect methods but uses fewer parameters. The lightweight DuDoCFNet predicts the most accurate $\mu$-maps among all the testing groups and has fewer parameters than CDI-Net (NMSE, 11.43$\%$ versus 12.19$\%$, p$<$0.001).

\begin{table} [htb!]
\caption{Quantitative evaluations of reconstructed AC SPECT images using predicted projections and $\mu$-maps on 250 testing cases. The best results are marked in \textbf{BOLD}.}
\label{tab:ac} 
\footnotesize
\centering
\resizebox{0.46\textwidth}{!}{
\begin{tabular}{ l | c | c | c || c }
\hline
\textbf{Methods}         & \textbf{NMSE($\boldsymbol{\%}$)}   & \textbf{SSIM}   & \textbf{PSNR}  & \textbf{P-values$^{a}$}   \Tstrut\Bstrut\\  
\hline  
Baseline Non-AC                        & $35.17 \pm 10.35$    & $.665 \pm .034$       & $24.10 \pm 1.82$        & $<$ 0.001      \Tstrut\Bstrut\\  
\hline \hline
U-Net \cite{ronneberger2015u}        & $6.55 \pm 1.88$      & $.863 \pm .021$       & $31.44 \pm 1.59$        & $<$ 0.001      \Tstrut\Bstrut\\  
\hline
DenseUNet \cite{huang2017densely}   & $6.27 \pm 1.80$      & $.864 \pm .020$       & $31.62 \pm 1.56$        & $<$ 0.001      \Tstrut\Bstrut\\  
\hline
AttenUNet \cite{oktay2018attention} & $6.18 \pm 1.84$      & $.864 \pm .021$       & $31.63 \pm 1.56$        & $<$ 0.001      \Tstrut\Bstrut\\  
\hline
DuDoSS \cite{chen2023dudoss}        & $6.12 \pm 1.83$      & $.865 \pm .021$       & $31.67 \pm 1.56$        & $<$ 0.001      \Tstrut\Bstrut\\  
\hline
TSP$\&$BDA-Net [Fig. \ref{fig:tsp}, \ref{fig:bda}]  & $5.38 \pm 1.73$      & $.875 \pm .021$       & $32.26 \pm 1.66$        & $<$ 0.001      \Tstrut\Bstrut\\  
\hline
CDI-Net \cite{chen2023cross}        & $5.26 \pm 1.61$      & $.876 \pm .021$       & $32.33 \pm 1.65$        & $<$ 0.001      \Tstrut\Bstrut\\  
\hline \hline
DuDoCFNet w/o TSP                 & $4.87 \pm 1.43$      & $.880 \pm .020$       & $32.67 \pm 1.63$        & $<$ 0.001      \Tstrut\Bstrut\\  
\hline
DuDoCFNet w/o BDA                 & $4.78 \pm 1.40$      & $.882 \pm .019$       & $32.73 \pm 1.61$        & $<$ 0.001      \Tstrut\Bstrut\\  
\hline 
DuDoCFNet w/o MLF                 & $4.75 \pm 1.32$      & $.882 \pm .019$       & $32.76 \pm 1.60$        & $<$ 0.001      \Tstrut\Bstrut\\  
\hline
\textbf{DuDoCFNet (Proposed)}           & $\mathbf{4.34 \pm 1.32}$    & $\mathbf{.889 \pm .019}$    & $\mathbf{33.07 \pm 1.67}$  & \textendash  \Tstrut\Bstrut\\  
\hline
\multicolumn{5}{l}{$^{a}$P-values of paired t-tests on NMSE between the current testing group and DuDoCFNet (Proposed).} \\
\end{tabular}
}
\end{table}

\subsection{Reconstructed AC SPECT Images}  
The predicted projections and $\mu$-maps are then incorporated into the ML-EM reconstruction (30 iterations) to produce AC SPECT images. TSP$\&$BDA-Net refers to the AC images reconstructed using predicted projections from TSP-Net and predicted $\mu$-maps from BDA-Net. Fig. \ref{fig:ac} shows the AC SPECT images. TSP$\&$BDA-Net outputs more accurate AC images than existing single-task learning methods. The AC images generated by DuDoCFNet are more consistent with the ground truth, compared to single-task learning methods, CDI-Net, and the ablation groups. Table \ref{tab:ac} lists the voxel-wise quantitative evaluations of the AC images. TSP$\&$BDA-Net outputs more accurate results than existing single-task learning methods. In addition, DuDoCFNet produces more accurate AC images than the other testing methods and the ablation groups (NMSE, 4.34$\%$ versus 5.26$\%$, p$<$0.001). This proves that DuDoCFNet significantly improves the final AC reconstruction accuracy.

\begin{table} [htb!]
\caption{Segment-wise quantitative evaluations of polar maps on 100 testing cases using APE, Correlation Coefficient (Corr. Coef.), and Coefficient of Determination (R$^{2}$). The best results are marked in \textbf{BOLD}.}
\label{tab:polar} 
\scriptsize
\centering
\resizebox{0.46\textwidth}{!}{
\begin{tabular}{ l | c | c | c | c }
\hline
\textbf{Methods}  & \textbf{APE($\boldsymbol{\%}$)}   & \textbf{P-values$^{a}$}   & \textbf{Corr. Coef.}  & \textbf{R$^{2}$}   \Tstrut\Bstrut\\  
\hline  
Baseline Non-AC                        & $16.99 \pm 13.11$    & $<$ 0.001      & 0.7239      & 0.5240       \Tstrut\Bstrut\\  
\hline 
U-Net \cite{ronneberger2015u}        & $13.63 \pm 7.97$     & $<$ 0.001      & 0.8891      & 0.7904       \Tstrut\Bstrut\\  
\hline
DenseUNet \cite{huang2017densely}   & $12.62 \pm 8.00$     & $<$ 0.001      & 0.9029      & 0.8151       \Tstrut\Bstrut\\  
\hline
AttenUNet \cite{oktay2018attention} & $12.52 \pm 7.96$     & $<$ 0.001      & 0.9030      & 0.8154       \Tstrut\Bstrut\\  
\hline
DuDoSS \cite{chen2023dudoss}        & $12.07 \pm 7.92$     & $<$ 0.001      & 0.9037      & 0.8167       \Tstrut\Bstrut\\  
\hline
TSP$\&$BDA-Net [Fig. \ref{fig:tsp}, \ref{fig:bda}]    & $9.13 \pm 7.04$      & $<$ 0.001      & 0.9093      & 0.8268       \Tstrut\Bstrut\\  
\hline
CDI-Net \cite{chen2023cross}        & $9.03 \pm 7.14$      & $<$ 0.001      & 0.9149      & 0.8370       \Tstrut\Bstrut\\  
\hline \hline
DuDoCFNet w/o TSP                 & $8.30 \pm 6.84$      & $<$ 0.001      & 0.9159      & 0.8388       \Tstrut\Bstrut\\  
\hline
DuDoCFNet w/o BDA                 & $8.25 \pm 6.27$      & $<$ 0.001      & 0.9262      & 0.8579       \Tstrut\Bstrut\\  
\hline
DuDoCFNet w/o MLF                 & $8.13 \pm 6.10$      & $<$ 0.001      & 0.9282      & 0.8615       \Tstrut\Bstrut\\  
\hline
\textbf{DuDoCFNet (Proposed)}           & $\mathbf{7.16 \pm 5.77}$     & \textendash       & \textbf{0.9315}       & \textbf{0.8677}      \Tstrut\Bstrut\\  
\hline
\multicolumn{5}{l}{$^{a}$P-values of paired t-tests on APE between the current testing group and DuDoCFNet (Proposed).} \\
\end{tabular}
}
\end{table}

\begin{figure}[htb!]
\centering
\includegraphics[width=0.48\textwidth]{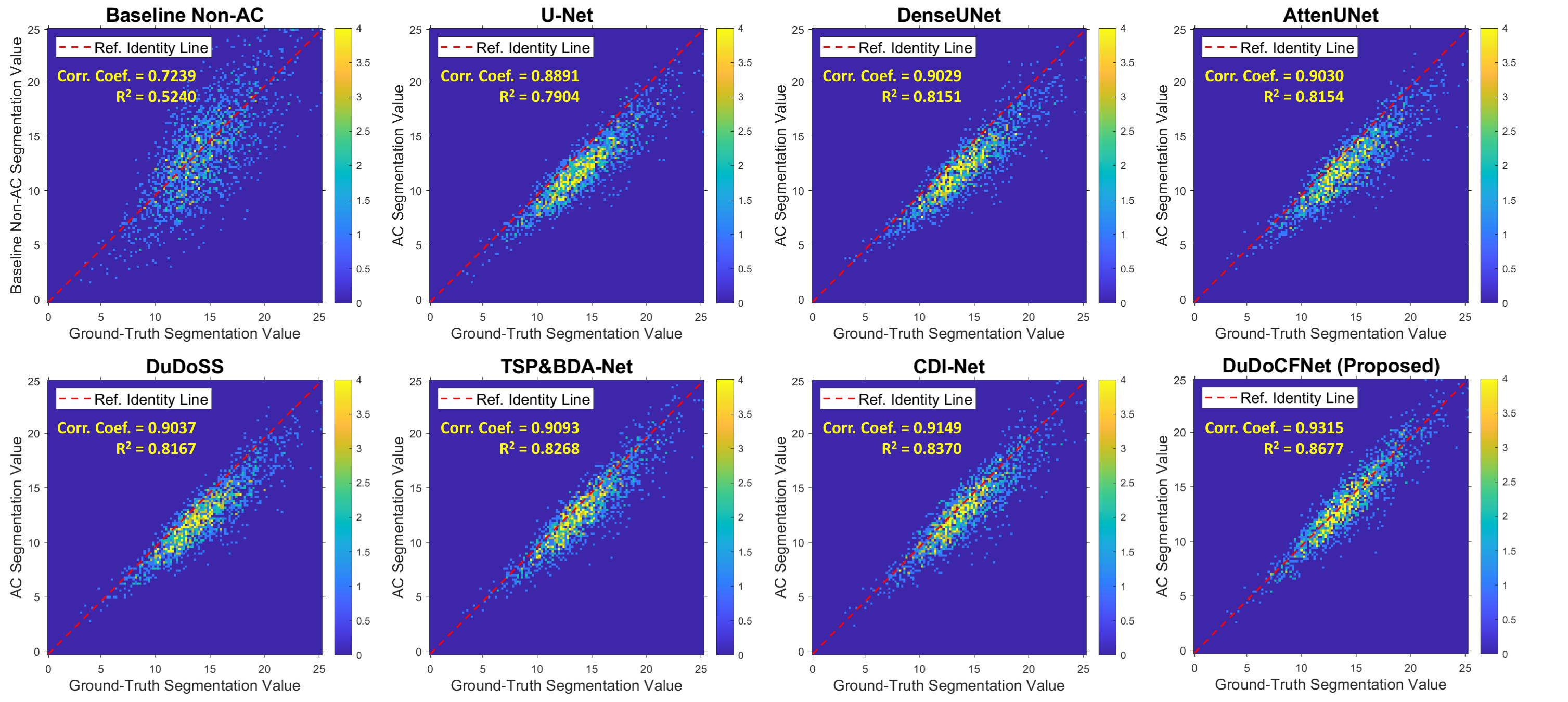}
\caption{Correlation maps of segment values between polar maps of the ground-truth and estimated AC SPECT images. Correlation Coefficients (Corr. Coef.) and Coefficients of Determination (R$^{2}$) are annotated.}
\label{fig:ac_polar_corr}
\end{figure}

The clinical standard 17-segment polar maps are generated from the AC SPECT images as shown in Fig. \ref{fig:ac_polar}. The TSP$\&$BDA-Net generates more accurate polar maps than existing single-task learning methods. In addition, the polar map by DuDoCFNet is the most consistent with the ground truth. Table \ref{tab:polar} presents the segment-wise quantification of the polar maps. The polar maps by DuDoCFNet demonstrate the lowest segment-wise errors (APE, 7.16$\%$ versus 9.03$\%$, p$<$0.001) and the highest Correlation Coefficient (Corr. Coef.) among all testing groups. The correlation maps of the segment-wise values are shown in Fig. \ref{fig:ac_polar_corr}. DuDoCFNet shows the most concentrated point distributions and the highest Corr. Coef. This further proves that DuDoCFNet produces superior AC reconstruction results. The polar map patterns by DuDoCFNet are highly consistent with the clinical ground truth, which could largely enhance the SPECT imaging accuracy and thus improve the clinical diagnosis of coronary artery diseases. 

\subsection{Impact of Iterations and Low-Dose Levels} 
\begin{figure}[htb!]
\centering
\includegraphics[width=0.48\textwidth]{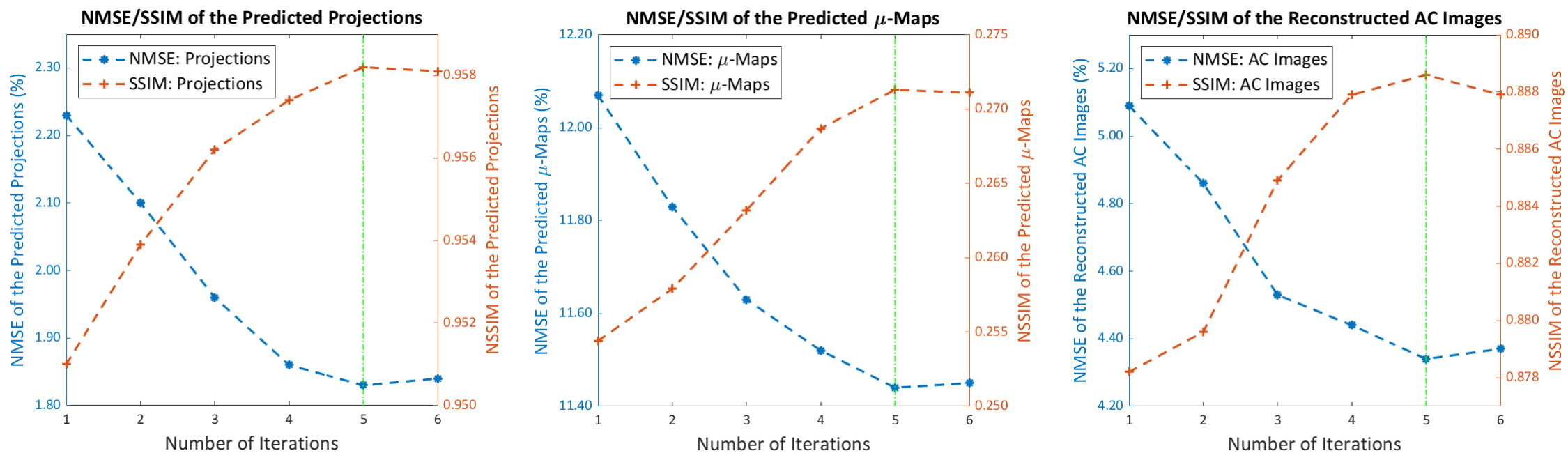}
\caption{Quantitative evaluations of DuDoCFNet with multiple iterations in predicted projections, $\mu$-maps, and AC images using NMSE and SSIM.}
\label{fig:iter}
\end{figure}

\begin{figure}[htb!]
\centering
\includegraphics[width=0.48\textwidth]{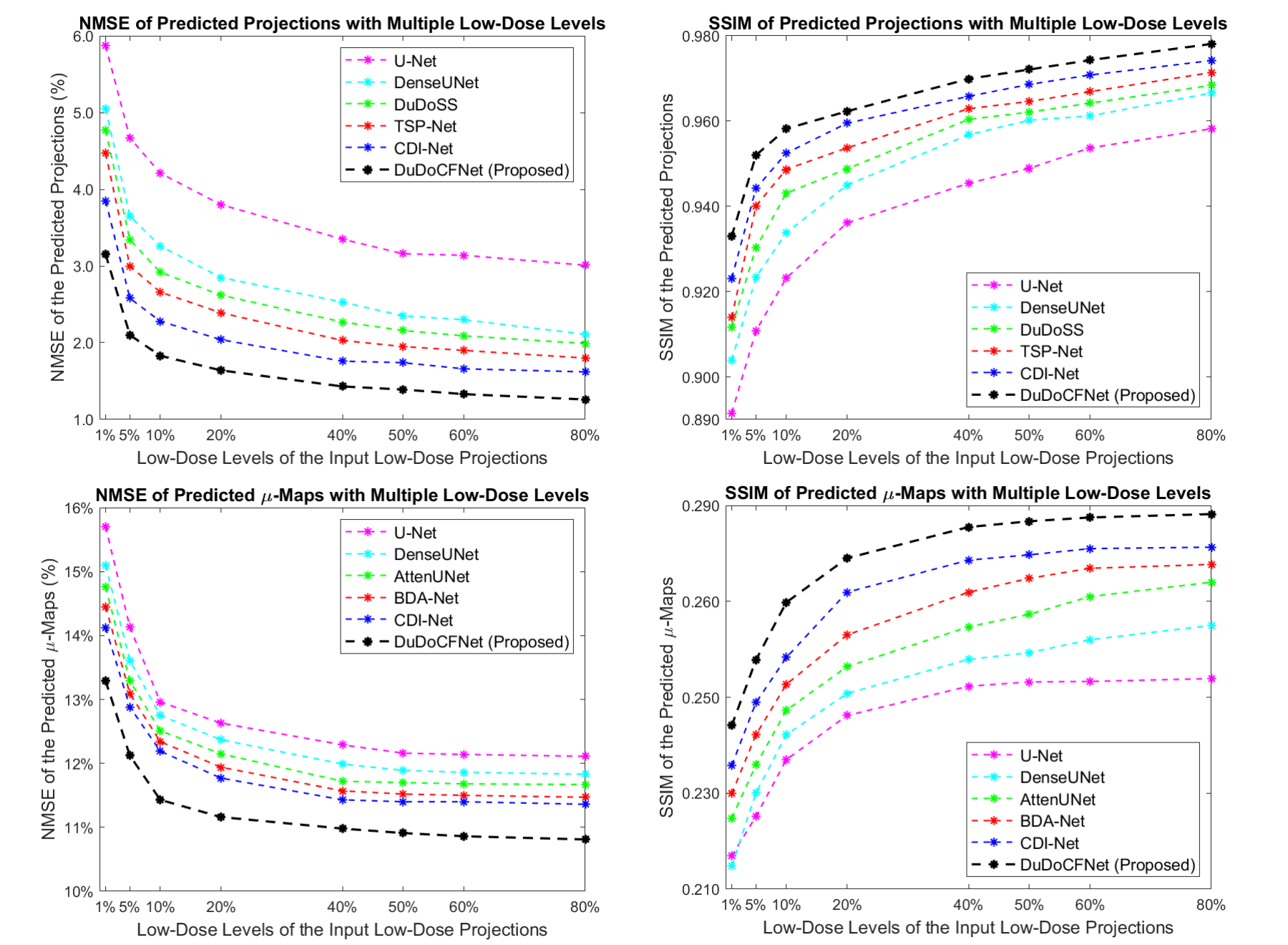}
\caption{Quantitative evaluations of multiple groups over 8 datasets with low-dose levels varying from 1$\%$ to 80$\%$ using NMSE and SSIM. }
\label{fig:dose}
\end{figure}

We further evaluated the performance of DuDoCFNet with different numbers of iterations as shown in Fig. \ref{fig:iter}. The average voxel-wise errors of the predicted projections, $\mu$-maps, and AC images by DuDoCFNet gradually decrease as the number of iterations increases, eventually converging at 5 iterations. In addition, we generated additional 7 datasets with varying LD levels ranging from 1$\%$ to 80$\%$. These datasets were used to evaluate DuDoCFNet and existing methods under different LD levels as shown in Fig. \ref{fig:dose}. It can be observed that DuDoCFNet exhibits consistently superior accuracy in estimating projections and $\mu$-maps under various dose settings.

\section{Discussion and Conclusion}
In this work, we propose DuDoCFNet, a multi-task learning method, for simultaneous LD denoising, LV reconstruction, and CT-free $\mu$-map generation of cardiac SPECT. Specifically, DuDoCFNet employs a dual-domain cascaded framework that enables cross-domain feature fusion. The MLF mechanism effectively connects and fuses the cross-modality image features at different spatial dimensions. The embedded CDF modules are utilized to adaptively adjust the channel-wise weights to enhance the cross-modality feature fusion. The TSP-Net and BDA-Net within DuDoCFNet share similar design motivations that utilize two-stage architectures for progressive coarse-to-fine estimations of projections or $\mu$-maps. Specifically, TSP-Net employs a U-Net-like structure to restore general structures and a non-downsampling module to recover finer details of the LA and LD projections. BDA-Net employs a shared encoder to predict a coarse $\mu$-map and the boundary, followed by a spatial co-attention module to generate a refined $\mu$-map with enhanced boundary accuracy. To minimize computational costs, DuDoCFNet is designed to be super lightweight by reducing the number of channels in its convolutional layers.

Our experiments based on clinical data demonstrate that DuDoCFNet predicts more accurate projections and $\mu$-maps than existing single-task learning methods. Additionally, DuDoCFNet has fewer network parameters but performs better than the previous multi-task learning framework CDI-Net, as indicated in Tables \ref{tab:proj} and \ref{tab:amap}. Ablation studies of DuDoCFNet further validate the impact of the proposed two-stage progressive coarse-to-fine estimation strategies and MLF mechanism on improving network performance. Moreover, we prove that DuDoCFNet's performance improves as the number of iterations increases and converges at 5 iterations, as illustrated in Fig. \ref{fig:iter}. Notably, even with only 1 iteration, DuDoCFNet still outperforms TSP-Net in predicting projections (2.23$\%$ versus 2.66$\%$, p $<$ 0.001) and BDA-Net in predicting $\mu$-maps (12.07$\%$ versus 12.34$\%$, p $<$ 0.001). This further proves that the simultaneous learning framework of DuDoCFNet improves the prediction accuracy of each interrelated task. In addition, we demonstrate that DuDoCFNet consistently exhibits superior performance under varying LD levels from 1$\%$ to 80$\%$, as presented in Fig. \ref{fig:dose}. Finally, we evaluate the reconstruction accuracy of DuDoCFNet in terms of the reconstructed AC SPECT images and the clinical standard 17-segment polar maps. The polar map patterns by DuDoCFNet are highly consistent with the ground truth, which could largely enhance the SPECT imaging accuracy and thus improve the clinical diagnostic capabilities for coronary artery diseases.

Our current work also has some potential limitations. First, DuDoCFNet employs an iterative framework that consists of cascaded TSP-Nets and BDA-Nets. Consequently, the computational costs associated with DuDoCFNet are relatively high compared to the single-task learning methods such as DenseUNet \cite{huang2017densely} and AttenUNet \cite{oktay2018attention}, particularly when multiple iterations are employed. However, even with just 1 iteration, DuDoCFNet still demonstrates promising performance. The inference time of DuDoCFNet is $<$ 1s even with high iterations, which is a reasonable speed for clinical practice. Second, in this study, DuDoCFNet is only tested using clinical data from cardiac SPECT-CT scanners. More validations of DuDoCFNet with diverse datasets from various tracers, scanners, organs, and imaging modalities could be conducted to comprehensively access DuDoCFNet's capabilities under different clinical scenarios. Last, the primary emphasis of this study lies in the methodology development and validation of DuDoCFNet. While we have generated and analyzed the standard 17-segment polar maps using clinical analysis tools, the clinical validation of DuDoCFNet remains insufficient and warrants further investigation and evaluation before being employed in clinical practice. 

Our work suggests some promising directions for future studies. First, DuDoCFNet indicates the significance of cross-modality feature fusion in enhancing prediction accuracy. As indicated in this study, the complementary information from emission and anatomical images enables more comprehensive feature representations and thus higher prediction accuracy. Hence, leveraging multi-modality features can potentially improve the performance of many medical imaging applications such as segmentation, registration, etc. Second, the multi-stage progressive learning strategy offers a viable solution for many complicated applications. As shown in our study, separating a complex learning task into several sections and utilizing specialized network modules for each individual section can enhance the overall performance. Last, this study showed that simultaneous learning of interrelated tasks can largely enhance the performance of each task. Therefore, further investigation is warranted for interrelated tasks such as denoising and registration\cite{chen2022dual}, which can be learned simultaneously. 


In conclusion, we propose DuDoCFNet for simultaneous LD denoising, LV reconstruction, and CT-free AC of cardiac SPECT. DuDoCFNet enables accurate and accelerated AC SPECT imaging while reducing hardware expenses and minimizing radiation exposure. The cascaded framework fuses the cross-domain and cross-modality image features for simultaneous learning. The two-stage progressive learning strategies improve estimation accuracy in both projection and image domains. Experiments with clinical data exhibit the superior performance of DuDoCFNet in predicting projections, generating $\mu$-maps, and AC reconstructions, compared to existing single- or multi-task learning methods. The clinical segment-wise evaluations using standard polar maps demonstrate that the AC reconstructions by DuDoCFNet are highly consistent with the clinical ground truth. This could largely enhance the SPECT MPI imaging accuracy and further improve the clinical diagnostic capabilities for coronary artery diseases. 

\bibliographystyle{IEEEtran}
\bibliography{refs}

\end{document}